%% file: main_ifac.tex
\documentclass{ifacconf}

\usepackage{filecontents}

\makeatletter
\let\old@ssect\@ssect 
\makeatother

\usepackage{natbib}
\usepackage{hyperref}

\makeatletter
\def\@ssect#1#2#3#4#5#6{%
	\NR@gettitle{#6}
	\old@ssect{#1}{#2}{#3}{#4}{#5}{#6}
}
\makeatother

\usepackage{dsfont,amssymb,amsmath,fancyhdr,mdframed}
\usepackage[dvipsnames]{xcolor}
\usepackage{comment}
\usepackage{dsfont}
\usepackage{graphicx}%
\usepackage{multicol,lipsum}
\usepackage{enumitem}
\usepackage{tikz}
\usepackage{hyperref}
\usepackage{subcaption}
\usepackage[prependcaption,colorinlistoftodos]{todonotes}

\usetikzlibrary{arrows.meta} 
\usetikzlibrary{fit, positioning}
\usetikzlibrary{calc}
\usetikzlibrary{decorations.pathmorphing} 

\tikzstyle{block} = [rectangle, minimum width=1cm, minimum height=1cm, text centered, draw=black]
\tikzstyle{tallblock} = [rectangle, minimum width=.5cm, minimum height=1cm, text centered, draw=black]
\tikzstyle{line} = [thick,-,>=stealth]
\tikzstyle{arrow} = [thick,->,>=stealth]
\tikzstyle{roundedblock} = [rectangle, minimum width=4cm, minimum height=2cm, text centered, draw=black, rounded corners=0.2cm]

\newcommand{\rezasay}[1]{\textcolor{DarkOrchid}{[\textsc{Reza:} #1]}}
\newcommand{\todoing}[1]{\todo[inline,color=green!20, linecolor=orange!250]{\small\emph{EB}: #1}}
\newcommand{\emiliosay}[1]{\todoing{#1}}
\newcommand{\R}{\mathbb{R}}
\newcommand{\N}{\mathbb{N}}
\newcommand{\mc}{\mathcal}
\newcommand{\bs}{\boldsymbol}
\newcommand{\col}{\mathrm{col}}
\newcommand{\buol}{\boldsymbol{u}^{\mathrm{OL}}}
\newcommand{\uol}{u^{\mathrm{OL}}}

\newcommand{\bu}{\boldsymbol{u}}
\newcommand{\Dx}{D^{\text{x}}}
\newcommand{\Du}{D^{\text{u}}}
\newcommand{\dx}{d^{\text{x}}}
\newcommand{\du}{d^{\text{u}}}
\newcommand{\X}{\mathbb{X}}
\newcommand{\U}{\mathbb{U}}
\newcommand{\red}[1]{\textcolor{red}{#1}}
\newcommand{\VI}{\mathrm{VI}}
\newcommand{\tsum}{\textstyle\sum}
\newcommand{\blkdiag}{\mathrm{blkdg}}
\newcommand{\blkmat}{\mathrm{blkmat}}

\newcommand{\row}{\mathrm{row}}

\newtheorem{definition}{Definition}
\newtheorem{lemma}{Lemma}
\newtheorem{assumption}{Assumption}

\newtheorem{proposition}{Proposition}
\newtheorem{proof}{Proof}
\newenvironment{myproof}{{\it Proof.}}{\hfill\qed}
\begin{document}
\begin{frontmatter}

\title{A Douglas-Rachford Splitting for solving Monotone affine Variational Inequalities in Linear-Quadratic Dynamic Games} 


\author[Erasmus]{Reza Rahimi Baghbadorani}    
\author[KTH]{Emilio Benenati}
\author[Delft]{Sergio Grammatico}              
\address[Erasmus]{Rotterdam School of Management, Erasmus University, The Netherlands (e-mail: rahimibaghbadorani@rsm.nl)}
\address[KTH]{Division of Decision and Control Systems, KTH Royal Institute of Technology, Sweden (e-mail: {benenati}@kth.se)}
\address[Delft]{Delft Center for Systems and Control, Delft University of Technology, The Netherlands (e-mail: s.grammatico@tudelft.nl)}

\begin{abstract}                
This paper considers constrained linear dynamic games with quadratic objective functions, which can be cast as affine variational inequalities. By leveraging the problem structure, we apply the Douglas-Rachford splitting, which generates a solution algorithm with linear convergence rate. The fast convergence of the method enables receding-horizon control architectures. Furthermore, we demonstrate that {the associated VI admits a closed-form solution within a neighborhood of the attractor, thus allowing for a further reduction in computation time.} Finally, we benchmark the proposed method via numerical experiments in an automated driving application.
\end{abstract}


\end{frontmatter}

\section{Introduction}\label{sec: intro}
Dynamic games provide a formal framework for analyzing multi-agent decision-making processes in dynamical systems, where the agents' objectives and constraints are coupled through both the system dynamics and their performance criteria \cite{basar_dynamic_1999, haurie2012games}. In such settings, each agent aims to optimize its own cost function while anticipating the actions of other agents, leading to strategic interactions that can be captured via various equilibrium concepts. Dynamic games are a useful model in robust control  \cite{theodor_output-feedback_1996}, as well as robotics \cite{wang_game-theoretic_2021, spica_real-time_2020}, electricity market control \cite{hall_2022} and supply chain strategic organization \cite{hall2024game}. These frameworks enable decentralized decision-making and the analysis of emergent behaviors in complex, multi-agent environments.\\
The desired solution concepts in dynamic games depend on the information structure assumed for the agents. One common approach is the open-loop Nash equilibrium (OL-NE), which defines a trajectory as a sequence of control inputs that is optimal for each agent, given the initial state and the input sequences chosen by the other agents \cite{monti2024feedback,sassano2021constructive}.
{The OL-NE problem can be cast as Variational Inequality (VI) \cite{benenati2024linear}, which can be solved by a plethora of algorithms \cite{facchinei2003finite}. This is especially relevant given the recent emergence of game-theoretic receding-horizon controllers \cite{lecleach_algames_2022, hall_stability_2024}, which define the control action at every time-step according to the solution to an OL-NE problem, thus requiring to complete its computation within the sampling time of the controller. This is especially difficult, as the complexity of the derived VI increases rapidly with the number of agents, the control horizon and the number of constraints. \\
In this work, we propose to leverage the affine structure of the VI resulting from a linear-quadratic game in order to define a specifically tailored solution algorithm based on the Douglas-Rachford iteration \cite{ferris1996operator}.
Let us justify our choice by illustrating the performance of some state-of-the-art VI solution algorithms, along with a simulation that highlights the contribution of this paper. 
Let us first define the VI problem as follows:
\begin{equation}\label{VI-main}
\text{find} \,\, u^* \in \mathcal{C} \quad \text{s.t.} \, \inf\limits_{u\in \mathcal{C}}\langle F(u^*), u - u^* \rangle \geq 0,
\end{equation}
where $\mathcal{C}\subseteq\R^n$ is a closed, convex set, and $F:\R^n\rightrightarrows \R^n$ is a (in general set-valued) operator. We assume that $F$ is monotone \cite{facchinei2003finite}, Lipschitz, and the solution set of \eqref{VI-main} is nonempty. Note that via the first-order optimality condition, we can rewrite \eqref{VI-main} as a monotone inclusion, $0 \in F(u) + \mathcal{N}_{\mathcal{C}}(u)$, where $\mathcal{N}_{\mathcal{C}}$ is the normal cone of the set \(\mathcal{C}\). This equivalence is beneficial and allows us to use operator splitting methods.\\
As shown in Section \ref{sec: OL-NE as VI}, {under some technical assumptions, a solution to a constrained linear-quadratic game can be found by solving a strongly monotone affine variational inequality problem} with linear constraints, i.e. $F = Mu + q$ and $\mathcal{C}: Du \leq d$ (where $M$ can be a non-symmetric matrix). Therefore, efficient numerical methods for solving affine VIs are essential in dynamic game equilibrium problems. \\
{\textbf{Illustrative example.} Consider the VI in \eqref{VI-main} with $\mathcal{C} = \mathbb{R}^n$ and $F = Mu$, where $M = I + S$, and $S = \begin{pmatrix}0 & -\omega \\ \omega & 0\end{pmatrix}$. 
The classical Forward-Backward (FB) splitting method generates the sequence
\begin{align*}
u^{k+1} = (I - \gamma(I + S)) u^k,
\end{align*}
whose convergence depends on the spectral radius of $T_{\mathrm{FB}} = I - \gamma(I + S)$, given by $\rho_{\mathrm{FB}} = 1 - \gamma(1 \pm i \omega)$. 
In contrast, the spectral radius of the  Douglas–Rachford (DR) splitting method, which we will discuss in greater detail in Section \ref{sec:convergence}, is given by
\[
\rho(T_{\mathrm{DR}}) = \tfrac{1}{2} \left( 1 + \frac{1 - \gamma j\omega}{1 + \gamma j\omega} 
\cdot \frac{1 - \gamma}{1 + \gamma} \right).
\]
For non-small stepsizes (e.g., $\gamma = 0.5$), the FB iteration exhibits non-decaying oscillations with $|\rho_{\mathrm{FB}}| = 1$, whereas DR contracts steadily toward equilibrium with $|\rho_{\mathrm{DR}}| \approx 0.52$.
As we will see in Section \ref{sec:convergence}, the DR takes advantage of the affine structure of the problem to split the operator that defines the VI in a symmetric and non-symmetric part. This enables the DR to effectively damp rotational modes, resulting in faster and more stable convergence for skew-dominated problems. \\
To demonstrate further effectiveness of the splitting method in a higher-dimensional VI problem, we refer readers to the appendix (Figure \ref{fig1}), where six widely used algorithms from the literature are compared.
}

{Motivated by exploiting the structure of linear VIs that arise from linear-quadratic dynamic games and the DR splitting to achieve faster convergence, our technical contributions are summarized as follows:}
\begin{itemize}
   \item {First, we define a strongly monotone affine VI whose solution, under some technical assumptions, is an infinite-horizon open-loop Nash equilibrium for the linear-quadratic dynamic game. We show that the solution to the problem is known in closed form when the initial state lies in a neighborhood of the origin.}
   \item We leverage the linearity of the VI to derive a tailored iterative solver based on the DR splitting method. Furthermore, we show linear convergence for the derived affine VI of OL-NE. 
   \item We adopt the aforementioned splitting method in computing the control input of a controller based on the receding-horizon solution to a VI derived from a linear-quadratic dynamic game. The test scenario is that of a set of autonomous vehicles, whose objective is to traverse a crossroad while maintaining a safe distance and speed.
\end{itemize}

\textit{\textbf{Notation.}} 
\begin{enumerate}[label=\alph*), itemsep = 0mm, topsep = 0mm, leftmargin = 7mm]
\item \textit{Matrices:} We denote as $I_n$ the identity matrix of size $n$ and as $0_{n,m}$ the zero matrix of size $n\times m$.  The subscripts are omitted when clear from context.  For a set of matrices of suitable dimensions indexed in $\mc I$, $(M_i)_{i\in\mc I}$, we denote as $\col(M_i)_{i\in\mc I}, \row(M_i)_{i\in\mc I}, \blkdiag(M_i)_{i\in\mc I} $ its column, row and diagonal stack, respectively. For a set of matrices of suitable dimensions indexed in $\mc I\times\mc J$, we denote the block matrix with $i,j$-th element $M_{ij}$ as $\blkmat(M_{ij})_{i\in\mc I, j\in\mc J}$ We denote $M^{-\top}=(M^{\top})^{-1}$ and the Kronecker product as $\otimes$. 
\item \textit{Euclidean spaces and operators:} Let $\mathcal{C}$ be the finite-dimensional real vector space with the standard inner product $\langle \cdot, \cdot \rangle$ Euclidean norm.  For $x\in\R^n$ and $Q\succ 0$, we denote $\|x\|_{Q} = \sqrt{x^{\top}Qx}$. The subscript is omitted when $Q=I$. We denote as $\pi_{\mathcal{C}}$ the metric projection onto set $\mathcal{C}$ ($\pi_{\mathcal{C}}(u) = \arg\min_{y \in \mathcal{C}} \|u - y\|$).  We define the normal cone of the set \( \mathcal{C} \) at the point \( u \in  \mathcal{C} \) as $\mathcal{N}_{\mathcal{C}} (u) = \{ g : g^\top u \geq g^\top y, \,\,  \forall y \in \mathcal{C} \}$. An operator $F$ is $L$-Lipschitz, if there is $L>0$ such that for all $u,y \in \mathcal{C}$ we have $\|F(u) - F(y)\| \leq L\|u-y\|$. The operator $F$ is (strongly) monotone if $\langle F(u) - F(y), u - y \rangle \geq  \mu \|u-y\|^2$ for all $u,y \in \mathcal{C}$ and some $\mu \geq 0 (>0)$. For a closed set $\mc C$, we denote its interior as $\mathrm{int}(\mc C)$. {We denote the VI problem \ref{VI-main} with the operator \( F \) and set \( \mathcal{C} \) as \( \mathrm{VI}(\mathcal{C}, F) \).} For an affine operator $F: x\mapsto Mx + q$, we denote the affine VI problem $\mathrm{AVI}(\mc C, M, q) := \VI(\mc C, F)$.
\item \textit{Real sequences:} We denote as $\mc S^n_T$ a sequence in $\R^n$ with length $T\in\N\cup \infty$. Let $w\in\mc S_T^n$. We denote as $w[t]$, with $t\in\{0,...,T-1\}$, its $t$-th element. If $w: \R^m \to \mc S^n_T$, that is, $w(x)$ is a sequence for all $x\in\R^m$, then we denote as $w[t|x]$ its $t$-th element. We treat finite sequences as column vectors, that is, if $w\in\mc S^n_T$, then $w=\col(w[t])_{t\in\{0,...,T-1\}}$.
\item \textit{Game theory:} We consider multi-agent systems, and we denote as $N$ the number of agents. We denote as $\mc I:=\{1,...,N\}$  and as $\mc I_{-i}:=\mc I\setminus\{i\}$. We denote in boldface the column stack over $\mc I$,  e.g. $\bs{w}=\col(w_i)_{i\in\mc I}$. We add the subscript $-i$ to denote the stack over $\mc I_{-i}$, e.g. $\bs{w}=\col(w_i)_{i\in\mc I_{-i}}$. If $w_i\in\mc S_\infty^N$ for all $i\in\mc I$ then, with slight abuse of notation, we denote $\bs{w}:=(w_i)_{i\in\mc I}$, $\bs{w}_{-i}:=(w_i)_{i\in\mc I_{-i}}$. 
\end{enumerate}
\section{Open-Loop Nash Equilibria as Solutions of a Variational Inequality}\label{sec: OL-NE as VI}
We consider a linear system with $N$ inputs, each controlled by a self-interest agent. We denote the sequence of control inputs of agent $i$ as $u_i \in \mc S_{\infty}^m$. The system dynamics is ruled by the difference equation
\begin{equation}\label{eq:dynamics}
    x[t+1] = Ax[t] + \sum_{i\in\mc I} B_i u_i[t].
\end{equation} 
We denote as $\phi(t, x_0, \bu)$ the solution at time $t$ of \eqref{eq:dynamics} with initial state $x_0$ and input $\bu$. We consider quadratic stage costs for all agents:
\begin{equation}
    \ell_i(x[t],u_i[t]):= \tfrac{1}{2}\|x[t]\|^2_{Q_i} + \tfrac{1}{2}\|u_i[t]\|^2_{R_i} \qquad \forall t
\end{equation}
such that, for all $i \in \mc I$, $Q_i=C_i^{\top}C_i \succcurlyeq 0$, $R_i = R_i^{\top}\succ 0$ and the pairs $(A,B_i)$ and $(A,C_i)$ are stabilizable and detectable, respectively. Furthermore, we consider linear time-invariant constraints: define the stage-feasible sets as
\begin{subequations}
\begin{align}
    \mathbb{U}_i(\bu_{-i}[t]):=\Large\{ u_i[t]: \textstyle\sum_{j\in \mc I} \Du_j u_j[t] + \du &\leq 0 \Large\} \label{eq:input_constraints} \\
    \mathbb{X}:= \{x[t]: \Dx x[t]+ \dx \leq 0 \}, \label{eq:state_constraints}
\end{align}
\end{subequations}
and the collective feasible input sequences with length $T\in\N\cup\{\infty\}$ as 
\begin{align}\label{eq:collective_constraints}
        \mc{U}_T(x_0):= \large\{ \bu:~~ &u_i[t] \in \U_i(\bu_{-i}[t]) &\forall i\in\mc I,t<T; \\
         &\phi(t,x_0,\bu) \in\X &\forall t<T\large\}.
\end{align}
We assume that the origin is strictly feasible, that is,
\begin{assumption}  \label{as:strict_feasibility}
    $0\in\mathrm{int}(\X)$; $0\in\mathrm{int}(\U_i(0))$, for all $i \in \mc I$.
\end{assumption}
Note that, if $T$ is finite, $\mc U_T(x_0)$ can be expressed as a system of affine inequalities by substituting \eqref{eq:dynamics} in \eqref{eq:collective_constraints} to eliminate $\phi$: 
\begin{align}
    \bu&\in\mc U_T(x_0) \iff D\bu + d \leq 0. \label{def-constraint5}
\end{align}
We define the (infinite-horizon) Open-Loop Nash equilibrium (OL-NE) $\buol$ as an input sequence such that the infinite-horizon objective
\begin{equation}
    J_i^\infty(u_i, \bs{u}_{-i}, x_0):= \sum_{t=0}^{\infty} \ell_i\big(\phi(t,x_0,\bu),u_i[t]\big)
\end{equation}
cannot be improved by unilateral modifications by any agent.
\begin{definition}[OL-NE] 
$\buol\in\mc{U}_{\infty}(x_0)$ is an OL-NE for the initial state $x_0$ if $\lim_{t\xrightarrow{}\infty} \phi(t,x_0,\buol)=0$ and
    \begin{equation}
        J_i^\infty(\uol_i, \buol_{-i}, x_0)\leq J_i^\infty(u_i, \buol_{-i}, x_0)
    \end{equation}
    for all $u_i$ such that $(u_i, \buol_{-i})\in\mc{U}_\infty(x_0)$.
\end{definition}
{The infinite-horizon problem is in general intractable. In the remainder of this section, we summarize the results of our previous work \cite{benenati2024linear}, where we identify an affine, finite-dimensional variational inequality (VI), whose solution are a truncation of the OL-NE under the following technical assumptions:}
\begin{assumption} \cite[Assumption 4.9]{monti2024feedback} \label{as:symplectic_matrix}
    $A$ is invertible and the matrix
    \begin{align*}
        H = \begin{bmatrix}
            A + \sum_{j\in\mc I}S_jA^{-\top}Q_j & \mathrm{row}(-S_jA^{-\top})_{j\in\mc I} \\
            \col(-A^{-\top}Q_j)_{j\in\mc I} & I_N \otimes A^{-\top}
        \end{bmatrix},
    \end{align*}
    where $S_i:=B_iR_i^{-1}B_i^{\top}$, has exactly $n$ eigenvalues with modulus strictly less than $1$. An $n$-dimensional invariant subspace of $H$ is complementary to
    \begin{equation*}
        \mathrm{Im}\left( \begin{bmatrix}0_{n\times Nn} \\ I_{Nn} \end{bmatrix} \right).
    \end{equation*}
\end{assumption}
We define $(P_i, K_i)_{i\in\mc I}$ as the solution to the coupled algebraic Riccati equations (ARE)
\begin{align}
    P_i &= Q_i + A^\top P_i(A +\tsum_{j\in\mc I}B_j K_j) \\
    K_i &= -R_i^{-1}B_i^{\top}P_i(A +\tsum_{j\in\mc I}B_j K_j);
\end{align}
such that the feedback gains $(K_i)_{i\in\mc I}$ stabilize the system in \eqref{eq:dynamics}. Such solution exists following \cite[Theorem 4.10]{monti2024feedback}. Furthermore, we define $(\hat{P}_i,\hat{K}_i)_{i\in\mc I}$ as the solutions to the (uncoupled) AREs
\begin{subequations}\label{eq:standard_are}
    \begin{align}
        \hat{P}_i &= \hat{Q}_i + \hat{A}_i^\top\hat{P}_i(\hat{A}_i+\hat{B}_i \hat{K}_i)\\
        \hat{K}_i &= -R_i^{-1}\hat{B}_i^{\top}\hat{P}_i(\hat{A}_i +\hat{B}_i \hat{K}_i),
    \end{align}
\end{subequations}
where, for each $i$
\begin{align}
    \begin{split}
        \hat{A}_i &:= \begin{bmatrix}
        A & \tsum_{j\neq i} B_j K_j \\
        0 & A + \tsum_{j\in\mc I} B_j K_j
    \end{bmatrix}, \\
    \hat{Q}_i &:= \blkdiag(Q_i, 0_{n\times n}),\\
    \hat{B}_i &:= \col(B_i, 0_{n\times m}),
    \end{split}
\end{align}
such that the feedback gains $\hat{K}_i$ stabilize the augmented system $(\hat{A}_i, \hat{B}_i)$ for each $i$. \\
Finally, let us introduce $\X_f$ as a forward-invariant, constraint-admissible set for the dynamics 
$$x[t+1]=(A + \tsum_{j\in\mc I} B_j K_j) x[t]. $$
Then, we can find a $T$-long truncation of an OL-NE sequence as the solution to the finite-horizon equilibrium problem
\begin{align} \label{eq:finite_hor_problem}
\begin{split}
\text{find $\bu^*\in\mc U(x_0)$ such that}\\ 
    \forall i: u_i^* \in \arg\min_{u_i} & ~~J_i(u_i, \bu^*, x_0) \\
    \text{s.t.}&~~ (u_i, \bu^*_{-i})\in\mc U_T(x_0), 
\end{split}
\end{align}
where
\begin{align} \label{eq:objective_fin_hor}
\begin{split}
 J_i(u_i, \bu^*, x_0)&:=\textstyle\sum_{t=0}^{T-1} \Big\{\ell(\phi(t, x_0, u_i, \bu^*_{-i}), u_i[t]) \Big\}\\ 
    &~~+\frac{1}{2}\left\| \begin{bmatrix}
        \phi(T, x_0, u_i, \bu^*_{-i})\\
        \phi(T, x_0, u_i^*,\bu^*_{-i})
    \end{bmatrix} \right\|^2_{\hat{P}_i}\\
\end{split}
\end{align}
as formalized next:
\begin{lemma}\cite[Theorem 1]{benenati2024linear}
    Let $\bu^*$ solve \eqref{eq:finite_hor_problem}. Let $x_T:=\phi(T, x_0, \bu^*)$ and let $x_T\in \mathbb{X}_f$.
    Then, the sequence defined as
    \begin{equation}
        \forall i: \begin{cases}
            u_i^*[t] & \text{if} ~t < T,\\
            K_i (A + \tsum_j B_jK_j)^{t-T} x_T & \text{if} ~t\geq T
        \end{cases}
    \end{equation}
    is an OL-NE for the initial state $x_0$.
\end{lemma}
We now develop a connection between the solutions to the problem in \eqref{eq:finite_hor_problem} and the solutions to a VI. Via straightforward calculation, we can find matrices $\Theta, (\Gamma_i)_{i\in\mc I}$ such that the dynamics in \eqref{eq:dynamics} is rewritten as
\begin{equation}
        \begin{bmatrix}
            x[1] \\
            ... \\
            x[T]
        \end{bmatrix} = \Theta x_0 + \sum_{i\in\mc I} \Gamma_i \begin{bmatrix}
            u_i[0] \\
            ... \\
            u_i[T-1]
        \end{bmatrix}.
\end{equation}
Then, the partial gradients of the functions in \eqref{eq:objective_fin_hor} with respect to the first argument read as 
\begin{equation} \label{eq:def_F}
    \begin{bmatrix}
        \nabla_{1} J_i(u_1, \bu, x_0)\\
        ...\\
        \nabla_{1} J_i(u_N, \bu, x_0)
    \end{bmatrix} = M\bu + q,
\end{equation}
    where
    \begin{align}\label{eq:def_VI_matrices}
    \begin{split}
        M&:= \blkdiag(\bar{R}_i)_{i\in\mc I} + \blkmat(\Gamma_i^{\top}\bar{Q}_i\Gamma_j)_{(i,j)\in\mc I^2}, \\
        q&:= \col(\Gamma_i^{\top}\bar{Q}_i\Theta x_0)_{i\in\mc I},\\
        \bar{R}_i&:= I_T\otimes R_i, \qquad \forall i\\
        \bar{Q}_i&:= \blkdiag(I_{T-1}\otimes Q_i, P_i), \qquad \forall i.
    \end{split}
    \end{align}
A sufficient condition for $\bu^*$ to solve \eqref{eq:finite_hor_problem} is that $\bu^*$ solves 
\begin{align}\label{eq:def_P2}
\begin{split}
\mathrm{AVI}(\mc{U}_T(x_0), M, q). 
\end{split}
\end{align}

\begin{lemma}\cite[Proposition 2]{benenati2024linear}\label{lem: OL-NE as VI}
If $\bu^*$ is a solution to the VI in \eqref{eq:def_P2}, then it is also a solution to the finite-horizon equilibrium problem in \eqref{eq:finite_hor_problem}.
\end{lemma}
\begin{figure}
    \centering
    \centering\resizebox{.7\columnwidth}{!}{\input{block_closed_GNE.tex}}
    \caption{Block scheme of the closed-loop dynamics with receding-horizon open-loop Nash equilibrium controller.}
    \label{fig:block_scheme}
\end{figure}
In Section \ref{sec:convergence}, we leverage the problem reformulation provided by Lemma \ref{lem: OL-NE as VI} to propose a fast solution algorithm to \eqref{eq:def_P2}. To further alleviate the computational burden, we show that the solution to \eqref{eq:def_P2} can be computed in closed-form when the initial state belongs to $\mathbb{X}_f$.
 \begin{proposition}\label{lem: inactive cons}
    Let $x_0\in \mathbb{X}_f$ and $M$ in \eqref{eq:def_VI_matrices} be such that $M+M^{\top}\succ 0$. Let for all $i$
    $$u^*_i[t] = K_i(A+\textstyle\sum_{i\in\mc I} B_iK_i)^tx_0.$$
    Then, the unique solution to the VI in \eqref{eq:def_P2} is the sequence
    $$\bu^*[0], ...,\bu^*[T-1].$$
\end{proposition}
\begin{myproof}
   From the definition of $\X_f$, $\bu^*$ is strictly feasible. We proceed by contradiction, and we assume there exists an input sequence $v_i$ such that
    \begin{equation}\label{eq:contradiction_assumption}
        J_i(v_i, \bu^*, x_0) < J_i(u_i^*,\bu^*, x_0).
    \end{equation}
    Define the sequences
     \begin{align}
     x^v[t] &= \phi(t, x_0, v_i, \bu_{-i}^*)\\
     x^*[t] &= (A+\textstyle\sum_{i\in\mc I} B_iK_i)^tx_0\\
    y[t] &=  \left(\hat{A}_i + \hat{B}_i \hat{K}_i\right)^t \begin{bmatrix}
            x^v[T] \\ x^*[T]
        \end{bmatrix} \label{eq:def_y} \\
        w_i[t]  &= \hat{K}_i y[t].
    \end{align}
    From the definition of $\hat{A}_i, \hat{B}_i$, we deduce the structure
    $$y[t] = \begin{bmatrix}
        z[t] \\ x^*[t]
    \end{bmatrix} $$
    where $z$ satisfies
    \begin{align}
        \begin{split}
            z[t+1] = Az[t] + B_i\hat{K}_i y[t] + \tsum_{j\neq i} B_j K_j x^*[t]
        \end{split}
    \end{align}
    Note that $K_j x^*[t] = \bu^*[t]$, $\hat{K}_iy[t] = w_i[t]$ and $z[0] = x^v[T]$, thus we can write
    \begin{equation}\label{eq:z_as_state_evol}
        z[t] = \phi(t, x^v[T], w_i, \bu^*_{-i}).
    \end{equation}
    Following \cite[Theorem 21.1]{hespanha_linear_2018} and the fact that $\hat{P}_i, \hat{K}_i$ solve the ARE in \eqref{eq:standard_are}, 
    \begin{align}\label{eq:inf_hor_objective}
    \begin{split}
        \frac{1}{2}\left\|\begin{bmatrix}
            x^v[T] \\ x^*[T]
        \end{bmatrix} \right\|^2_{\hat{P}_i} =&\textstyle\sum_{t=0}^\infty \tfrac{1}{2}\|y[t]\|^2_{\hat{Q}_i} + \tfrac{1}{2}\|w_i[t]\|^2_{\hat{R}_i} \\
       =& \textstyle\sum_{t=0}^\infty \tfrac{1}{2}\|z[t]\|^2_{Q_i} + \tfrac{1}{2}\|w_i[t]\|^2_{R_i} \\ 
       =& \textstyle\sum_{t=0}^\infty \ell_i(z[t], w_i[t]),
    \end{split}
    \end{align}
    where we used the definition of $\hat{Q}_i, \hat{R}_i$. Denote for all $i$
    \begin{align} \label{eq:v_inf}
        &v_i^{\infty}[t] :=\begin{cases}
            v_i[t] & \text{if}~t<T \\
            w_i[t-T] & \text{if}~t\geq T.
        \end{cases} 
    \end{align}
    After substituting \eqref{eq:inf_hor_objective} in \eqref{eq:finite_hor_problem},
    \begin{align} 
    \begin{split}
        J_i(v_i, \bu^*,x_0) =& \tsum_{t=0}^{T-1} \ell(\phi(t, x_0, v_i, \bu^*_{-i}), v_i[t] ) \\
        +&\tsum_{\tau=0}^{\infty} \ell(z[t], w_i[t]).
    \end{split}
    \end{align}
We substitute \eqref{eq:z_as_state_evol} and \eqref{eq:v_inf}
\begin{align}
       J_i(v_i, \bu^*,x_0) =& \tsum_{t=0}^{T-1} \ell(\phi(t, x_0, v_i, \bu^*_{-i}), v_i[t] ) \\
        +&\tsum_{\tau=0}^{\infty} \ell(\phi(t, x^v[T], w_i, \bu^*_{-i}), w_i[t])\\
        =& \tsum_{t=0}^{\infty} \ell(\phi(t, x_0, v_i^{\infty}, \bu^*_{-i}), v^\infty_i[t] ) \\
        =& J_i^\infty(v_i^\infty, \bu^*_{-i}, x_0),
\end{align}
     Similarly, one can show
    \begin{equation}
        J_i({u}_i^*, {\bu}^*,x_0) = J_i^\infty(u^*_i, \bu_{-i}^*, x_0).
    \end{equation}
    Thus, \eqref{eq:contradiction_assumption} implies 
    $$J_i^\infty(v_i^\infty, \bu^*_{-i},x_0) < J_i^\infty( u^*_{i}, \bu^*_{-i}, x_0),$$
    which contradicts the fact that $\bu^*$ is an infinite-horizon OL-NE for the unconstrained system \cite[Prop. 1]{benenati2024linear}. Therefore, $\bu^*$ solves \eqref{eq:finite_hor_problem}. Because ${\bu}^*$ is a strictly feasible solution, the optimality conditions of \eqref{eq:finite_hor_problem} lead to
    \begin{equation}\label{eq:gradient_is_zero}
        \nabla_1 J_i({u}_i^*, {\bu}^*, x_0)  = 0, \quad \forall ~i
    \end{equation}
    From \eqref{eq:def_F}, we obtain
    \begin{equation}
         M\bu^* + q = 0.
    \end{equation}
     From the definition of VI in \eqref{VI-main}, we conclude that $\bu^*$ is a solution to the problem in \eqref{eq:def_P2}. The statement follows then by the uniqueness of the solution \cite[Prop. 2.3.2, Th. 2.3.3]{facchinei2003finite}.
    \end{myproof}
    
    With the receding-horizon control application illustrated in Figure \ref{fig:block_scheme}, in the remainder of this paper, we explore a Douglas-Rachford (DR) splitting algorithm that exploits the linear structure of the VI. 

\section{Algorithm and Convergence}\label{sec:convergence}
In this section, we first present a Douglas-Rachford splitting-like method for solving an affine variational inequality. Then, based on the results from the previous section, we tailor the proposed algorithm to the affine VI that defines $\mc P_2(x_0)$ in \eqref{eq:def_P2}. To this end, let us consider a general affine VI problem $\mathrm{AVI}(\mathcal{C}, M, q)$ and denote with $M_1, M_2, H$ matrices such that $M = M_1 + M_2$, and
\begin{align}\label{eq:conditions_convergence}
\begin{split}
        M_1 = M_1^{\top} &\succcurlyeq 0, \, M_2 \succ 0, \\
        H = H^{\top} &\succ 0.
\end{split}
\end{align}
Consider the following Douglas-Rachford splitting-like method \cite[Eqq. 30, 31]{ferris1996operator} to solve $\mathrm{AVI}(\mathcal{C}, M, q), \, k \in \mathbb{N}$:
\begin{subequations}\label{DR-affineVI}
\begin{align}
    y^{k} &= \mathrm{sol}(\mathcal{C}, H+M_1, q+(M_2 - H)u^k), \label{DR-affineVI-1}\\
    u^{k+1} &= (H+M_2)^{-1}\left(H(2\lambda_k y^{k} + (1-2\lambda_k)u^k) + M_2u^k\right), \label{DR-affineVI-2}
\end{align}
\end{subequations}
where $\mathrm{sol}(\mathcal{C}, M, q)$ denotes the solution to $\mathrm{AVI}(\mathcal{C}, M, q)$. At first glance, the iteration in \eqref{DR-affineVI} is not effective, as the step in \eqref{DR-affineVI-1} requires itself the solution of an affine VI. However, note that $H+M_1$ is a symmetric operator, differently from $M$. Therefore, the step in \eqref{DR-affineVI-1} is equivalent to solving an optimization problem \cite[\S 1.3.1]{facchinei2003finite}, and specifically a quadratic program (QP) in the case when $\mc C$ is a polyhedron, which can be solved very efficiently \cite{osqp}. The iteration method in \eqref{DR-affineVI} is well-defined under the condition in \eqref{eq:conditions_convergence}, as the \textrm{AVI} in \eqref{DR-affineVI-1} admits a unique solution, and $H+M_2$ is invertible. Linear convergence of the iteration in \eqref{DR-affineVI} to the solution of $\mathrm{VI}(\mathcal{C}, M, q)$ is provided in the following lemma.
\begin{lemma}[{Linear convergence \cite{ferris1996operator}}]\label{li-conv}
     The sequence $(u_k)_{k\in\N}$ generated by the iteration in \eqref{DR-affineVI} converges linearly \cite[Eq. 5.8]{bauschke_convex_2017} to the solution of $\mathrm{AVI}(\mathcal{C}, M, q)$ where $\lambda \in (0,1]$, $M = M_1 + M_2$ and $M_1$, $M_2$, and $H$ are matrices chosen according to \eqref{eq:conditions_convergence}.
\end{lemma}
\begin{myproof}
The result follows directly from Case 2 in \cite[Proposition 6]{ferris1996operator} and \cite[Proposition 8]{ferris1996operator}.  
\end{myproof}

We also refer interested readers to \cite{giselsson2017tight,moursi1805douglas}, where the authors provide different convergence guarantees for different kinds of splitting in inclusion problems, which can be extended to solving VI.\\
{We choose $H, M_1, M_2$ as follows:
\begin{subequations}\label{eq:example_matrix_splitting}
\begin{align}
H &= (1-\gamma)(M + M^\top) + \varepsilon I, \\
M_1 &= \gamma(M + M^\top), \label{eq:example_matrix_splitting:1} \\
M_2 &=  (M - M^\top) + (1-\gamma)(M + M^\top), \label{eq:example_matrix_splitting:2}
\end{align}
\end{subequations}
with $\gamma\in(0,1)$, and we recall that $M$ is defined in \eqref{eq:def_VI_matrices}.
This splitting choice satisfies the conditions in \eqref{eq:conditions_convergence} when $M\succ 0$ and it thus results in a linearly convergent Douglas-Rachford iteration according to Lemma \ref{li-conv}. It is designed so that, in the particular case $M=M^\top$ and with the choice $\lambda=\frac{1}{2}$,  \eqref{DR-affineVI} becomes
\begin{equation}
    u^{k+1} =  \mathrm{sol}(\mathcal{C}, \varepsilon I + M, q - \varepsilon u^k),
\end{equation}
which is equivalent to
\begin{equation}
    u^{k+1} = \arg\min_{u\in\mc C} \frac{1}{2} \|u + (M+\varepsilon I)^{-1}(q - \varepsilon u^k) \|^2_{M+\varepsilon I}.
\end{equation}
The latter, for $\varepsilon \approx 0$, resembles a full-step towards the unconstrained solution of \eqref{VI-main}, namely, $-M^{-1}q$, projected onto the constraint set. }
\section{Numerical Experiments}\label{sec: simulation}
We test the proposed Douglas-Rachford algorithm {with $H=I$, $\lambda_k = 0.5~ \forall k$, {and $M_1$, $M_2$ chosen as in \eqref{eq:example_matrix_splitting}} } on the receding-horizon control architecture in Figure \ref{fig:block_scheme}, applied to the control of autonomous vehicles driving through a crossroad. The vehicles traverse the intersection in the sequence of their arrival. Each vehicle aims to maintain a safe distance from the preceding vehicle intersecting their path, while matching its velocity. If no preceding vehicle intersects their path,  the agent is called a \emph{leading vehicle}, and its objective is to maintain a reference speed. We denote the set of leading vehicles as $\mc{L}$ and the preceding vehicle intersecting the path of $i$ as $\chi(i)$. {We simulate $N=15$ vehicles, with directions
\begin{align*}
    \{&\mathrm{NS}, \mathrm{ES}, \mathrm{WE}, 
    \mathrm{NW}, \mathrm{WN}, \mathrm{WN},
    \mathrm{WS}, \mathrm{NE}, \\
    &\mathrm{NE},
    \mathrm{EW}, \mathrm{NS}, \mathrm{ES}, \mathrm{WS}, \mathrm{SW}, \mathrm{WE}\}
\end{align*}
where \textrm{NS} denotes ``from North to South'' and so on.} For example: Agent 1 is a leading agent; Agent 2 must maintain a safe distance from Agent 1 because the paths \textrm{NS} and \textrm{ES} intersect, thus $\chi(2)=1$; Agent 4 (\textrm{NW}) must maintain a safe distance from Agent 1 (\textrm{NS}), because the path of agent 4 does not intersect with the ones of agents 2,3, thus $\chi(4)=1$. \\
We define $x=\col(x_i)_{i\in\mc I}$, where the state of each vehicle $x_i$ is
\begin{equation}
    x_i = \begin{cases} v^{\text{ref}} - v_i & \text{if}~i\in\mc L\\
        \begin{bmatrix}
            p_{\chi(i)} - p_i - d_i \\
            v_{\chi(i)} - v_i
    \end{bmatrix}  & \text{if}~i\notin\mc L.
    \end{cases}
\end{equation}
\begin{figure*}[!h]
    \centering
    \captionsetup{justification=centering}
    \subfloat[Agents 1--4 approach the intersection.]{\label{sub-fig1}\includegraphics[scale=0.4]{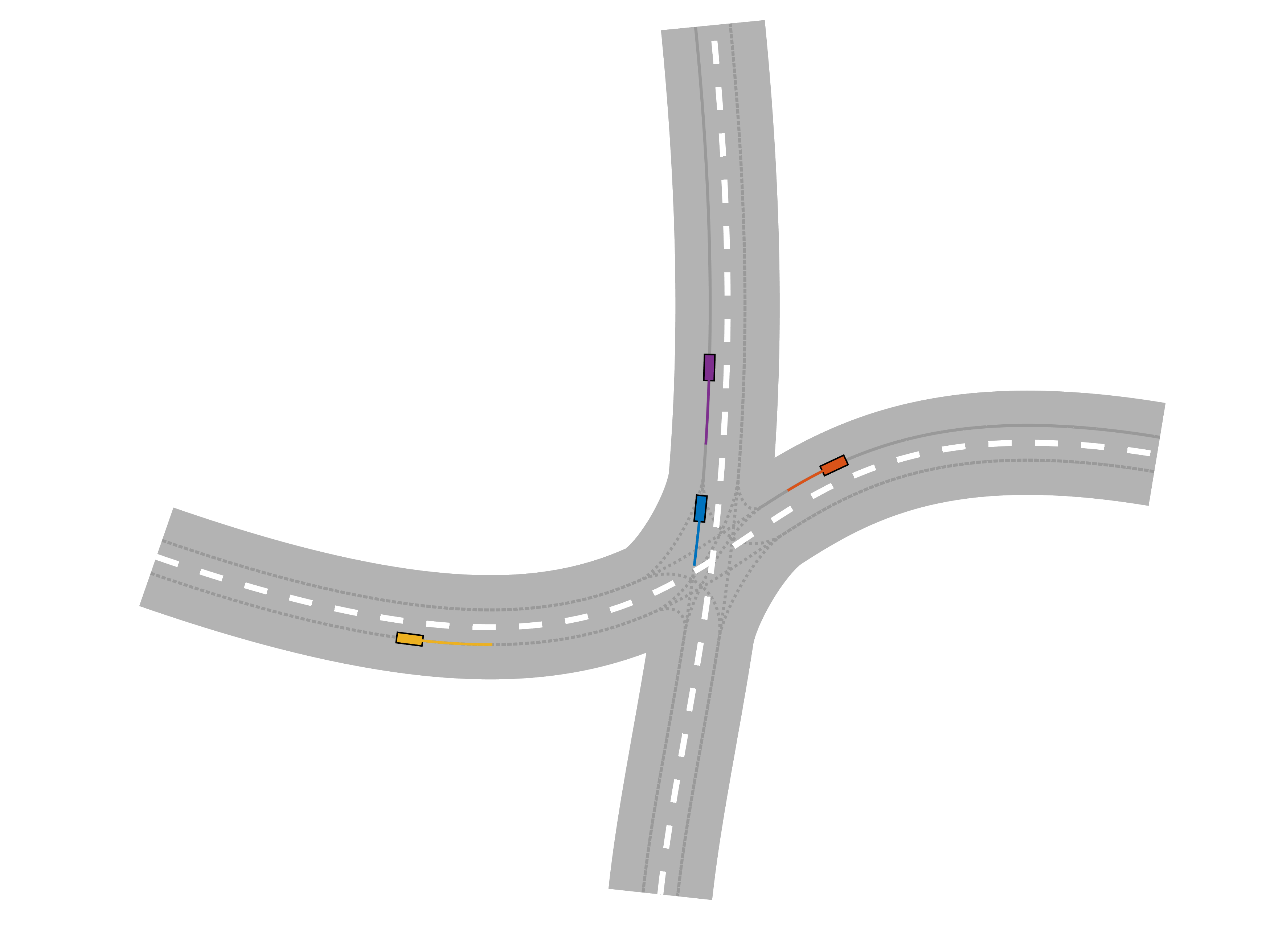}}
    \hspace{-2mm}
    \subfloat[Agent 1 (Blue) crosses the intersection first. Agents 2 (orange), and 4 (violet) cross next, simultaneously.]{\label{sub-fig2}\includegraphics[scale=0.4]{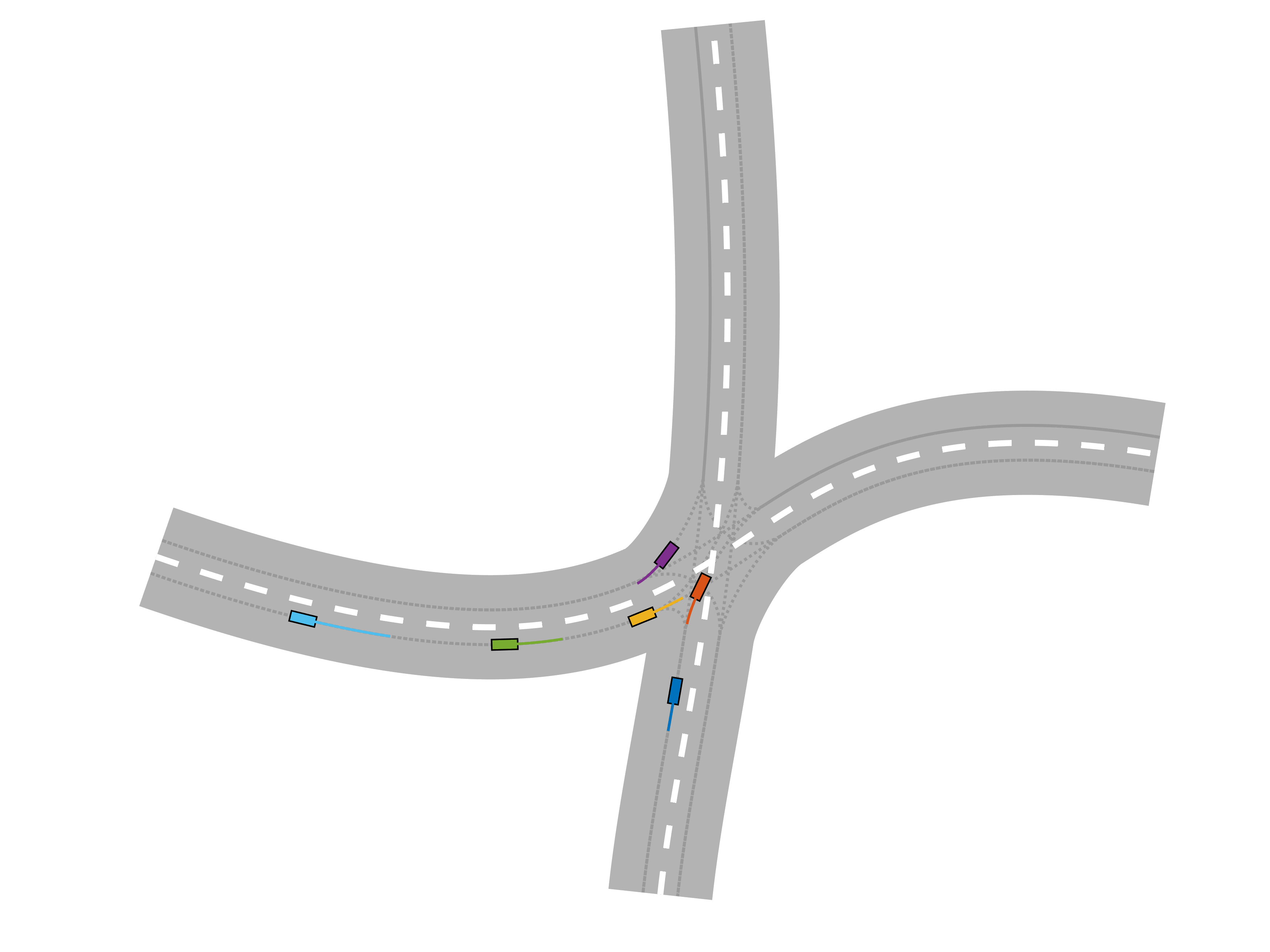}
    \hspace{-2mm}}
    \subfloat[Agent 3 (yellow) crosses the intersection, followed by Agent 5 (green).]{\label{sub-fig3}\includegraphics[scale=0.4]{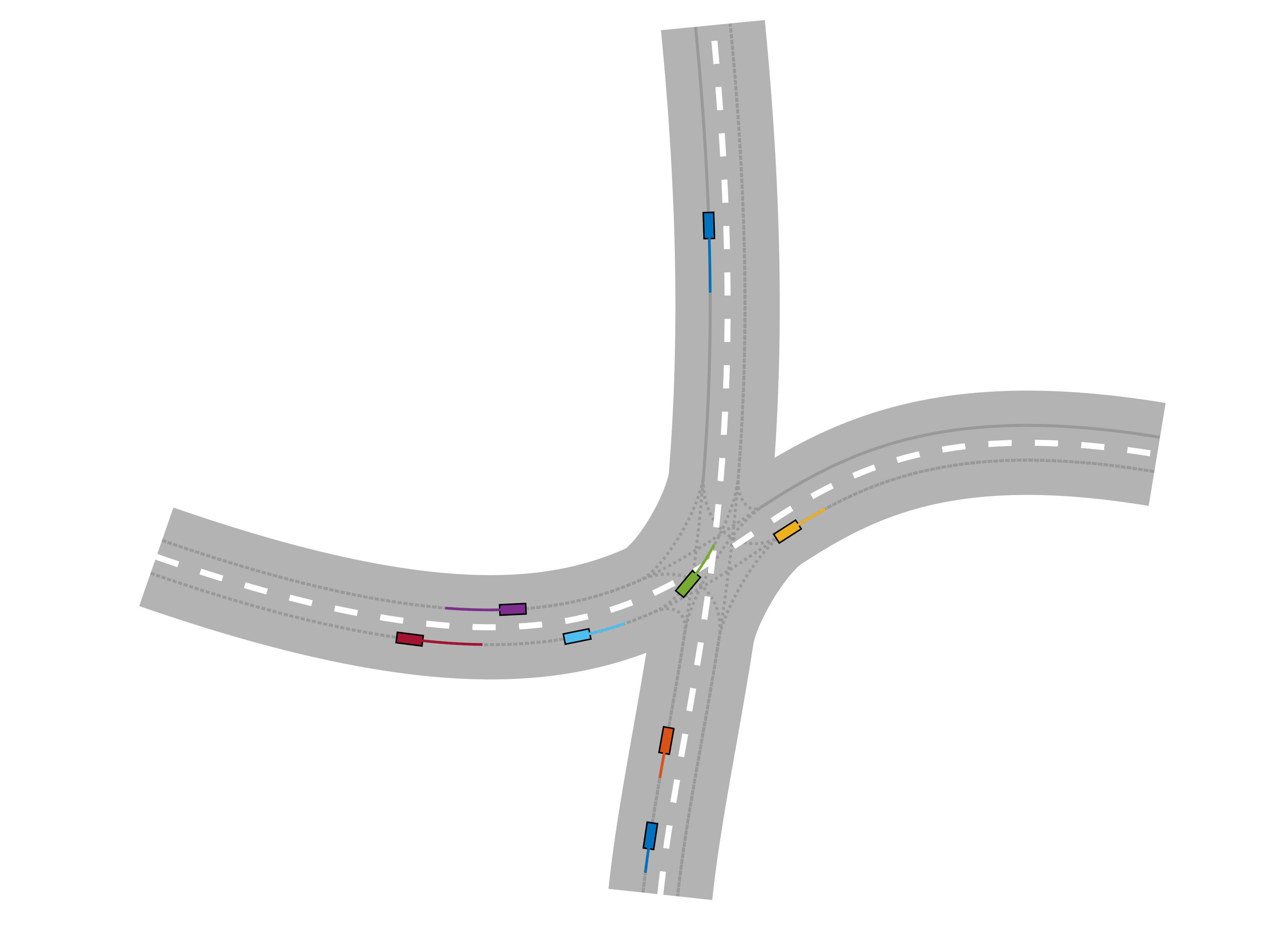}}\\
    \vspace{2mm}
        \subfloat[Agent 6 (cyan) crosses the intersection. Agents 7--9 approach.]{\label{sub-fig4}\includegraphics[scale=0.4]{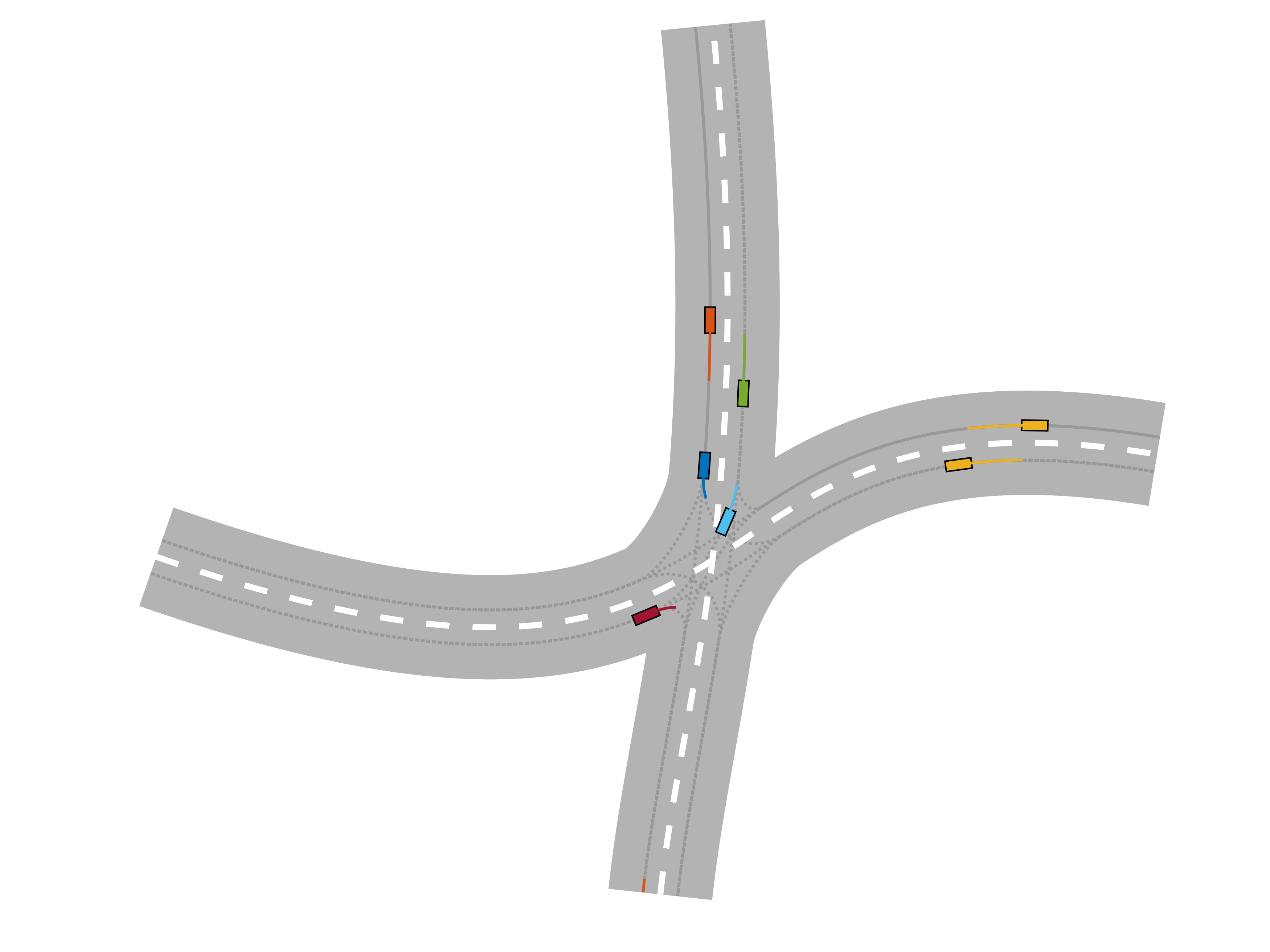}}
    \hspace{-2mm}
    \subfloat[Agents 7 (red) and 8 (blue) cross the intersection simultaneously, followed by agent 9 (orange).]{\label{sub-fig5}\includegraphics[scale=0.4]{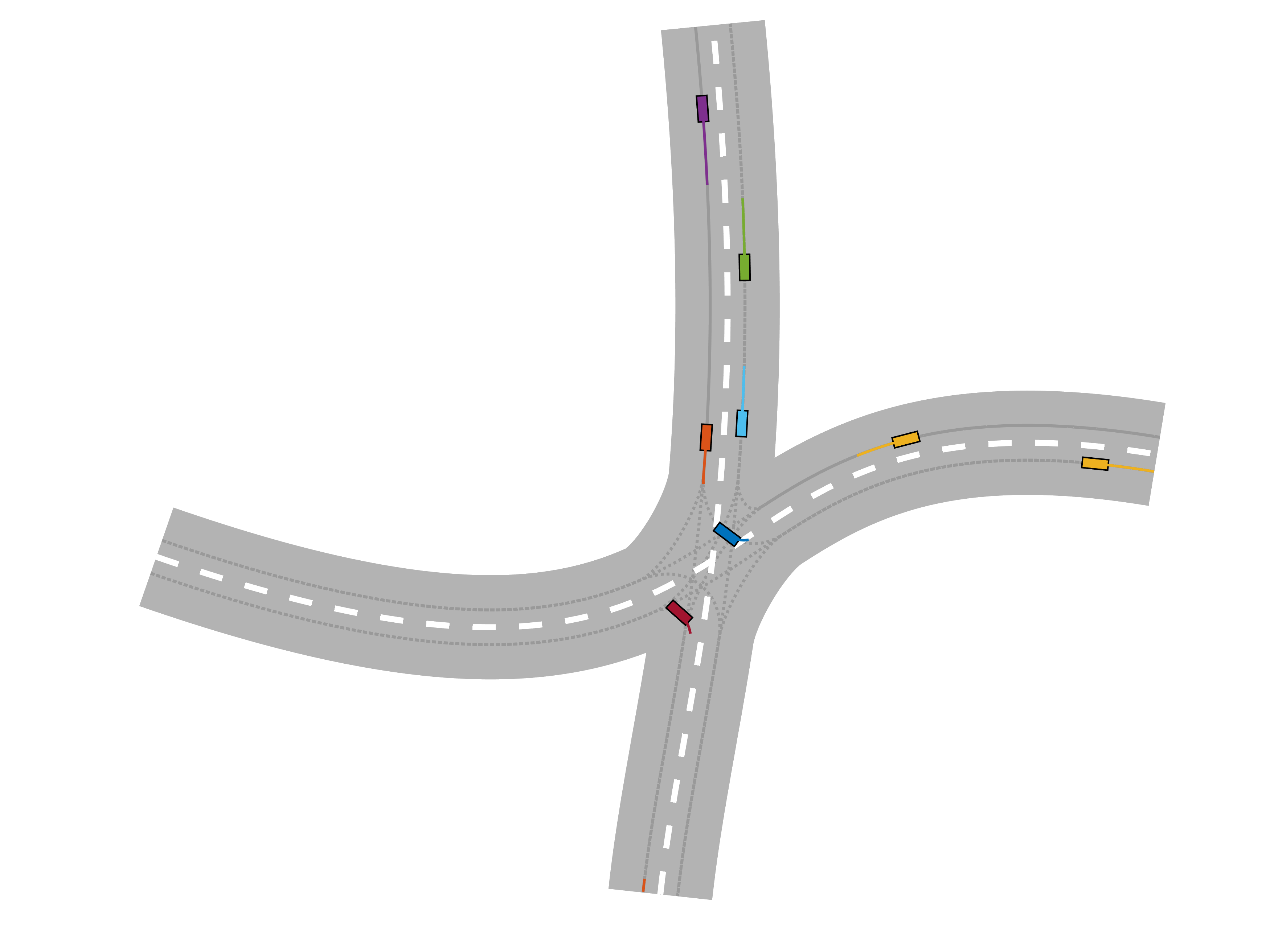}
    \hspace{-2mm}}
    \subfloat[Agent 10 (yellow) crosses the intersection, followed by Agent 11 (purple). Agents 12 (green), 13 (cyan), 14 (red) approach the intersection.]{\label{sub-fig6}\includegraphics[scale=0.4]{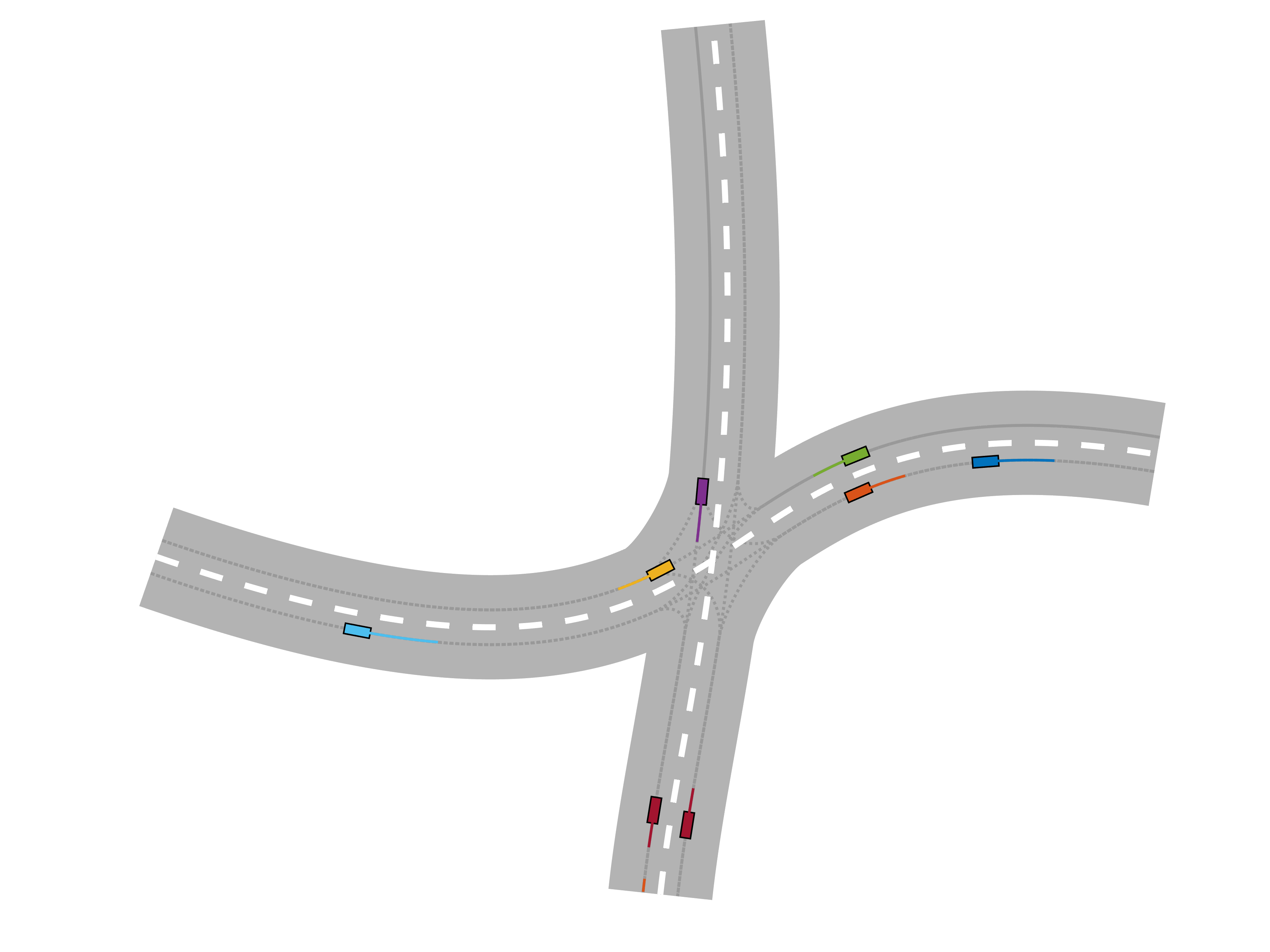}}\\
    \vspace{2mm}
        \subfloat[Agents 11 (purple) and 12 (green) cross the intersection, followed simultaneously by Agents 13 (cyan) and 14 (red).]{\label{sub-fig7}\includegraphics[scale=0.4]{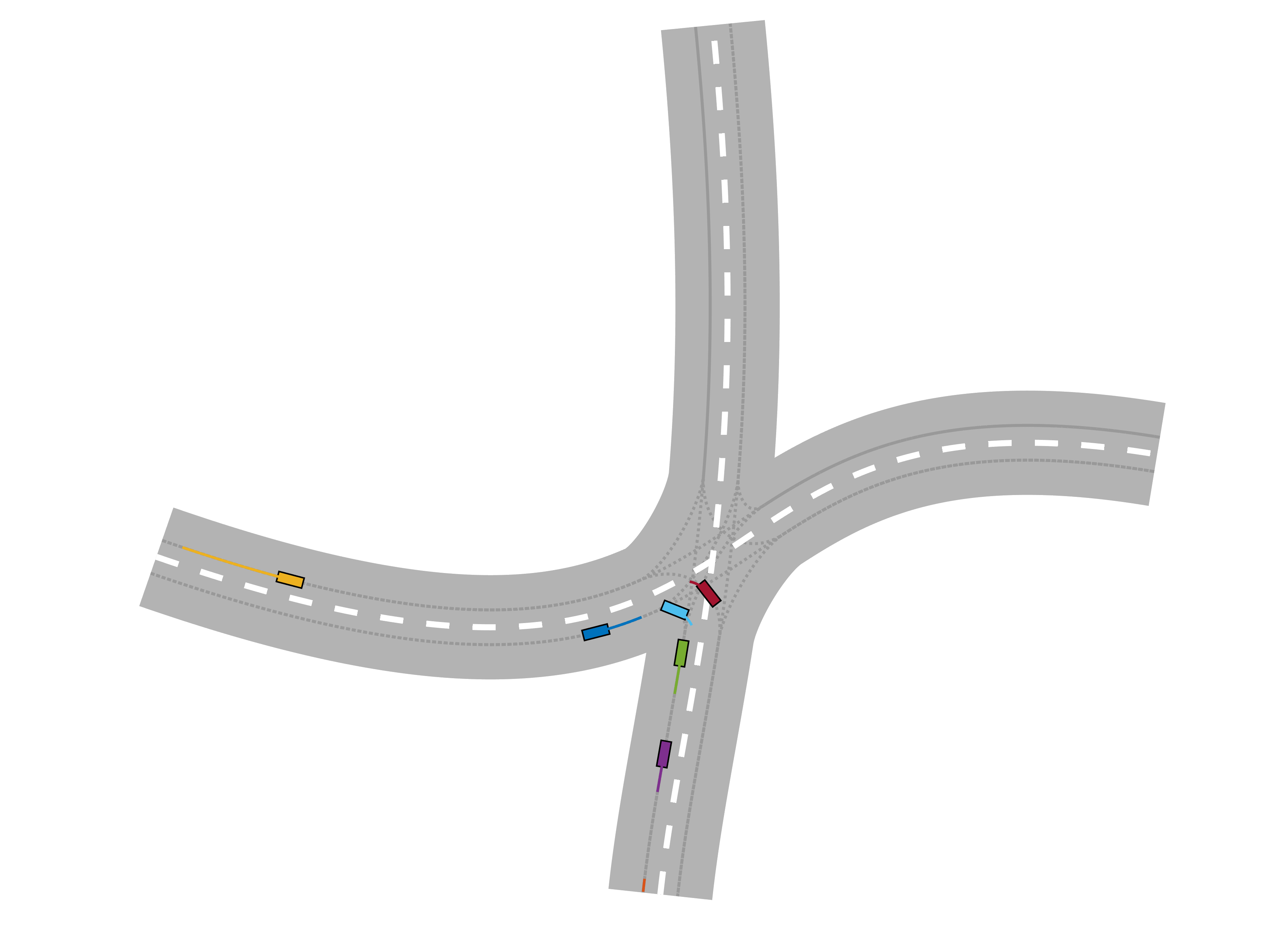}}
    \hspace{-2mm}
    \subfloat[Agent 15 (blue) crosses \\ the intersection last.]{\label{sub-fig8}\includegraphics[scale=0.4]{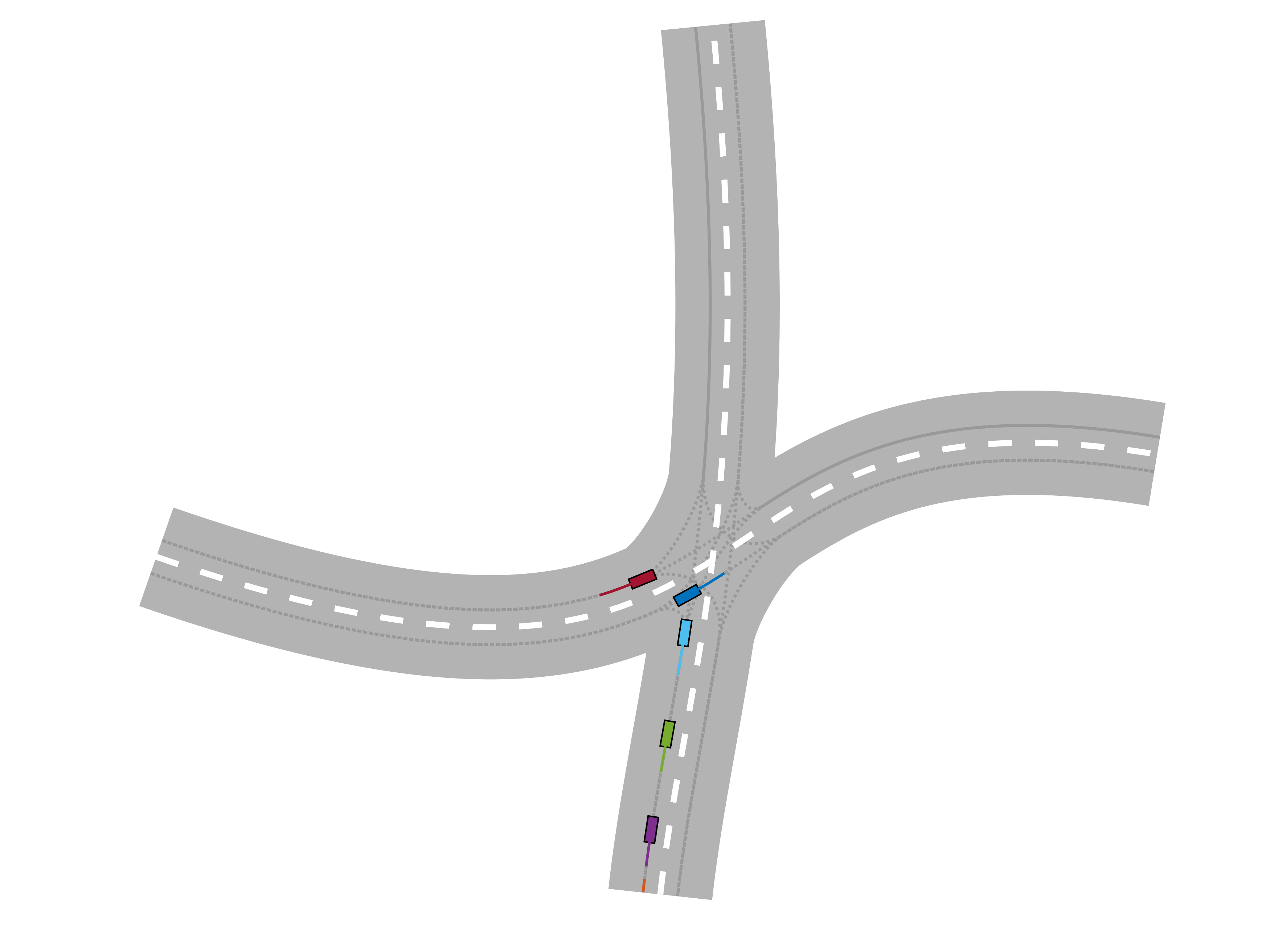}
    \hspace{-2mm}}
    \subfloat[The intersection is cleared.]{\label{sub-fig9}\includegraphics[scale=0.4]{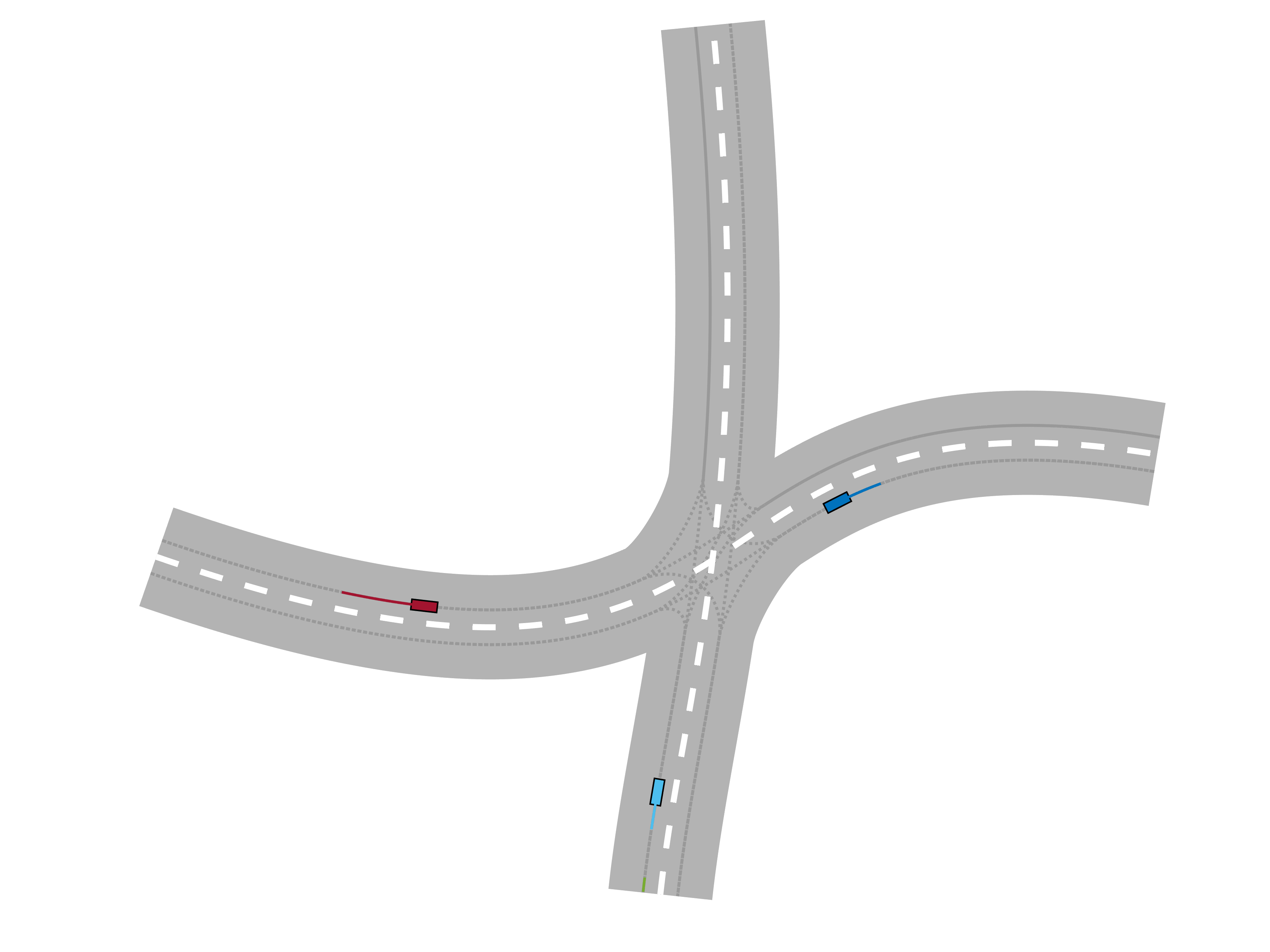}}
        \caption{{Simulated scenario. The full animation is available at \href{https://github.com/bemilio/CDC2025_DR_OLNE/blob/aa6a7633c8d4964b2ec475ff0e964d75041d4d2f/Simulations/vehicle_animation.mp4}{\texttt{https://tinyurl.com/444xr94f}.}}}
    \label{fig:frame_simulation}
\end{figure*}
In the latter, $p_i$ denotes the progress (in meters) of vehicle $i$ starting from the entrance point to the intersection, $d_i$ is the desired distance from $\chi(i)$, and $v_i$ denotes the speed. Each vehicle is modelled as a double integrator, discretized with rate $\tau_\text{s}=0.1s$. The matrices defining \eqref{eq:dynamics} are
\begin{align}
    \begin{split}
        A &= \blkdiag(A_i)_{i\in\mc I}, ~\text{where}~ A_i = \begin{cases}
            1 & \text{if} ~i\in\mc L, \\
            \begin{bmatrix}
                1 & \tau_{\text{s}} \\ 0 &1 
            \end{bmatrix} &\text{if} ~i\notin\mc L
        \end{cases} \\
        B_i &= \col(B_{ij})_{j\in\mc I}, ~\text{where}  \\
        B_{ij} &= \begin{cases}
            \begin{bmatrix}
            \tau_\text{s}^2/2 & \tau_\text{s} 
        \end{bmatrix}^\top &\text{if}~ i=\chi(j)\\
        -\begin{bmatrix}
            \tau_\text{s}^2/2 & \tau_\text{s} 
        \end{bmatrix}^\top &\text{if}~ i=j ~\text{and}~ i\notin\mc L \\
        -\tau_\text{s} &\text{if}~ i=j ~\text{and}~ i\in\mc L \\
        0_{1\times 2}  & \text{else}.
        \end{cases}
    \end{split}
\end{align}
In order to satisfy the stabilizability assumption, each agent applies the pre-stabilizing local controller $\bar{K}_i$ such that
\begin{equation*}
    \bar{K}_ix =-0.1\cdot \mathds{1}^\top x_i.
\end{equation*}
The constraints include the safety distance and the speed constraints, as well as an input box constraint:
\begin{align*}
    p_{\chi(i)} - p_i \geq d_{\text{min}},\\
    v^{\text{min}} \leq v_i \leq v^{\text{max}},\\
    u^{\text{min}} \leq u_i \leq u^{\text{max}}. 
\end{align*}
The state and input weight matrices are $Q_i=I_n$, $R_i = 1$. We observe in Figure \ref{fig:simulation_result} that all vehicles achieve the desired reference speed and distance, while satisfying the constraints. An animation of the simulated scenario is available at the provided link, where it is observed that the vehicles safely complete the maneuver. {We illustrate some extracted frames of the animation in Figure \ref{fig:frame_simulation}.} At each time-step, we apply the Douglas-Rachford iteration in \eqref{DR-affineVI}, warm-started at the ``shifted'' input sequence 
\begin{align*}
    \tilde{u}_i = (u^*_i[1], ..., u^*_i[T-1], K_ix^*_T), \qquad \forall i\in\mc I,
\end{align*}
where $\bu^*$ is the solution to \eqref{eq:def_P2}, where the initial state $x_0$  is substituted with the current state of the system, and $x^*_T = \phi(T, x[t], \bu^*(t))$. 
 {Figures \ref{fig:simulation_result_convergence} and \ref{fig:time_convergence} illustrate that the proposed method is significantly more computationally efficient than the Forward-Backward splitting method \cite{facchinei2003finite} and the ADMM-inspired method proposed in \cite{min_admm-iclqg_2025}, both in terms of the number of convergence iterations and the computational time. We note that the speed-up due to the warm-start is not available for the ADMM method. This is because the ADMM algorithm disregards the warm-start and, as a first step, moves the solution estimate to the one of a regularized unconstrained game. As shown, the number of iterations required to converge to the solution decreases with the evolution of the system, and we speculate that this is due to the system state converging towards the set $\mathbb{X}_f$, where the warm-start coincides with the solution to \eqref{eq:def_P2}, in view of {Proposition} \ref{lem: inactive cons}.} We consider that the algorithm has converged when \( r_k \leq 10^{-3} \), where \( r_k \) is the natural residual at iteration \( k \) of the iterative method \eqref{DR-affineVI} \cite[\S 6.2.1]{facchinei2003finite}.
\begin{figure}[!h]
    \centering
    \includegraphics[width=.8\linewidth]{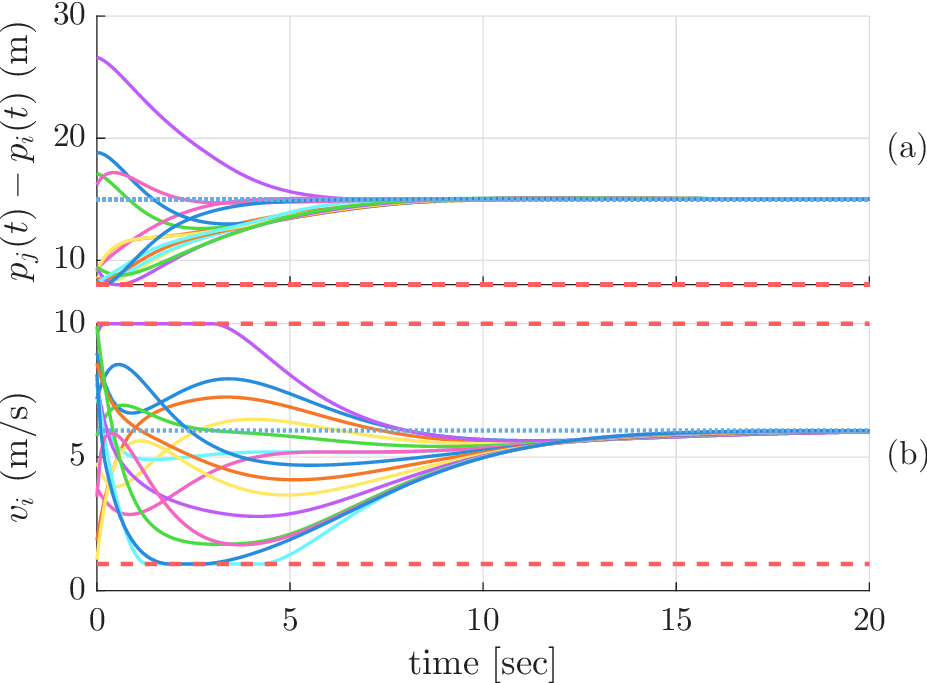}
    \caption{(a): Distance between $\chi(i)$ and $i$. (b): Velocity of each agent. The dotted lines denote the reference values, and the red dashed lines denote the constraints. }
    \label{fig:simulation_result}
\end{figure}
\begin{figure}[!h]
    \centering
    \includegraphics[width=.8\linewidth]{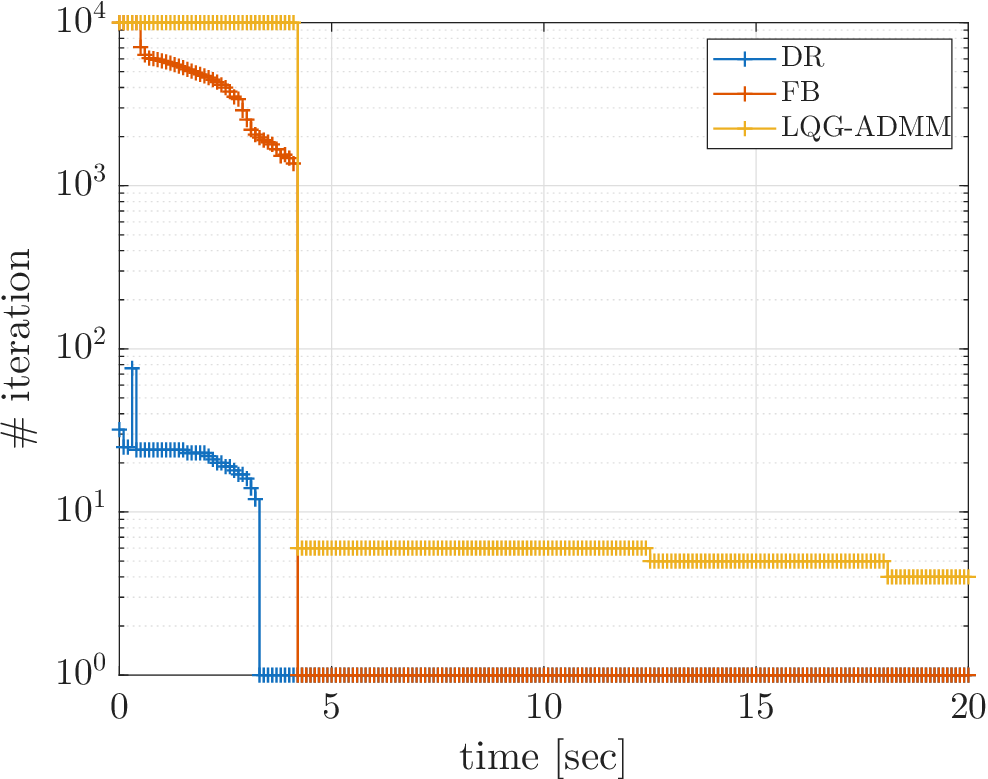}
    \caption{{Number of iterations to achieve convergence of the VI solution algorithm.}} 
    \label{fig:simulation_result_convergence}
\end{figure}
\begin{figure}[!h]
    \centering
    \includegraphics[width=.8\linewidth]{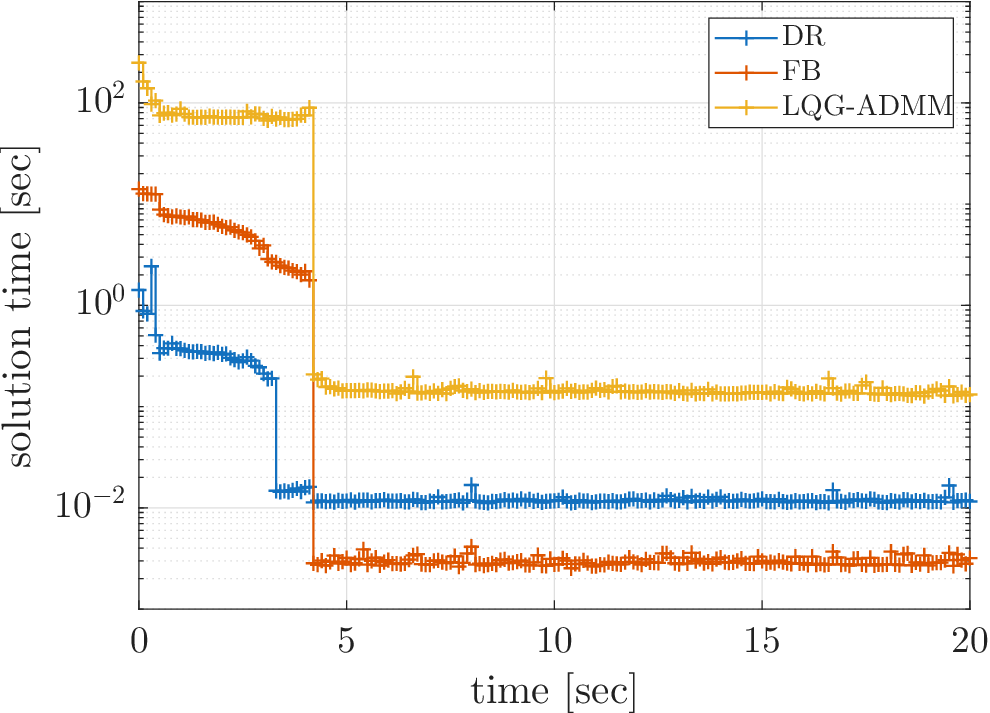}
    \caption{{Computational time to achieve convergence of the VI solution algorithm.}} 
    \label{fig:time_convergence}
\end{figure}
\section{Conclusion}\label{sec: conclusion}
For linear-quadratic dynamic games, an open-loop Nash equilibrium is the solution to an affine variational inequality, if the terminal state belongs to a forward-invariant, constraint-admissible set for the infinite-horizon unconstrained Nash equilibrium. The Douglas-Rachford splitting-like method is particularly suited for this affine variational inequality application, as it shows a remarkable convergence speed. Fast convergence to the solution enables the adoption of receding-horizon game-theoretic control architectures. Additionally, considering that the Nash equilibrium is known in closed-form near the equilibrium attractor of the dynamic game, {the computational effort required to solve the associated VI is further reduced by employing an informed warm-start.} Future work will investigate other applications of real-time control for multi-agent robotic systems based on receding-horizon variational inequalities.

\appendix \label{appendix}
\section{Comparison of iterative algorithms for affine variational inequalities}
{For a higher dimensional VI problem, Figure \ref{fig1} presents a comparison of the residuals for six widely used algorithms from the literature: (i) Forward-Backward splitting (FB) \cite{nemirovskij1983problem}, (ii) Extragradient descent (EXGD) \cite{malitsky2014extragradient} (iii) Nesterov's accelerated gradient descent (NAGD) \cite{nesterov2006solving} (iv) Projected reflected gradient descent (PRGD) \cite{malitsky2015projected}, (v) adaptive Golden ratio (aGraal) \cite{malitsky2020golden}, and (vi) Operator splitting methods (DR) \cite{facchinei2003finite} (additional details on these algorithms and their update rules are available in the next subsection). These algorithms are applied to 10 randomly generated strongly monotone affine VIs, with a non-symmetric matrix $M \in \mathbb{R}^{n \times n}$, a vector $q \in \mathbb{R}^n$, and linear constraints defined by $D \in \mathbb{R}^{m \times n}$ and $d \in \mathbb{R}^{m}$, where $n = 100$ and $m = 20$. As illustrated in Figure \ref{fig1}, the splitting method demonstrates superior performance compared to other methods.}
\begin{figure}[!h]
      \centering
           \includegraphics[width=0.4\textwidth]{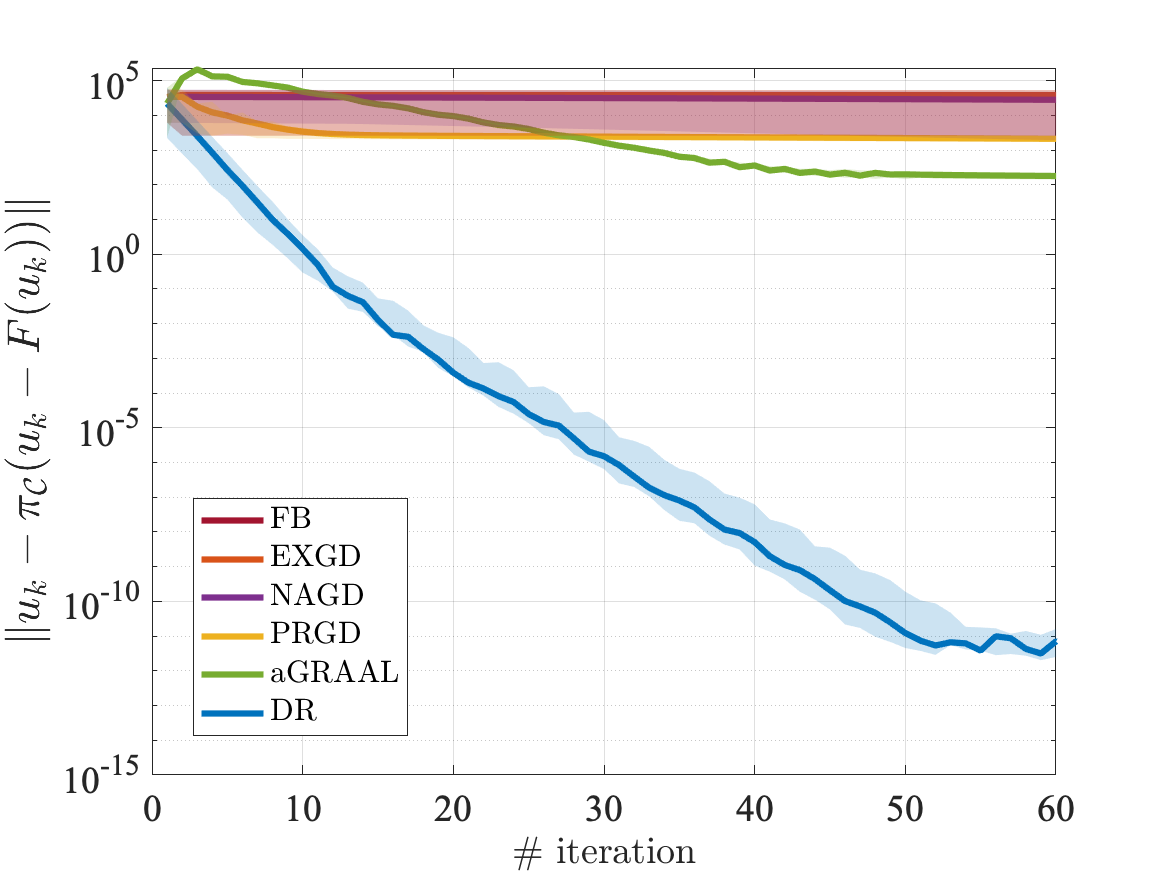}
            \caption{Comparison of residuals for different state-of-the-art VI solution methods.} 
            \label{fig1}
\end{figure}
\section{State-of-the-art algorithms for solving (strongly)~monotone variational inequalities} \label{algorithm review}
We review some recent and closely related existing methods for solving \eqref{VI-main}.

\textbf{(i)~Forward-Backward splitting (FB) \cite{nemirovskij1983problem}:}  
\begin{align*}
    u^{k+1} &= \pi_{\mathcal{C}}(u^k - \lambda F(u^k)),
\end{align*}
where \(\lambda\) is the stepsize. This method guarantees convergence for strongly monotone operators (with constant \(\mu\)) and Lipschitz operators (with constant \(L\)) when \(\lambda \in (0, 2\mu/L^2)\).\\
\textbf{(ii)~Extragradient Descent ({EXGD}) \cite{malitsky2014extragradient}:}
\begin{align*}
    y^k &= \pi_{\mathcal{C}}(u^k - \lambda F(u^k)), \\
    u^{k+1} &= \pi_{\mathcal{C}}(u^k - \lambda F(y^k)),
\end{align*}
with \(\lambda\) as the stepsize. Unlike FB, this method ensures convergence for Lipschitz operators (with constant \(L\)) when \(\lambda \in (0, 1/L)\). \\
\textbf{(iii)~Nesterov's accelerated gradient descent ({NAGD}) \cite{nesterov2006solving}:}
\begin{align*}
    u^k &= \arg\max_{u \in \mathcal{C}} \, \sum_{i=0}^k \lambda_i \left[\langle F(y^i), y^i - u \rangle - \frac{\mu}{2} \|u - y^i\|^2\right], \\
    y^{k+1} &= \arg\max_{u \in \mathcal{C}} \, \langle F(u^k), u^k - u \rangle - \frac{\beta}{2} \|u - u^k\|^2,
\end{align*}
where \(\lambda_{k+1} = \frac{\mu}{L} \sum_{i=0}^k \lambda_i\) is the stepsize at iteration \(k+1\). This method guarantees convergence for strongly monotone operators (with constant \(\mu\)) and \(\beta = L\), where \(L\) is the Lipschitz constant of the operator \(F\).\\
\textbf{(iv)~Projected Reflected Gradient Descent ({PRGD}) \cite{malitsky2015projected}:}  
\begin{align*}
    u^{k+1} &= \pi_{\mathcal{C}}(u^k - \lambda F(2u^k - u^{k-1})),
\end{align*}
where \(\lambda\) is the stepsize. This method guarantees convergence for a Lipschitz operator (with constant \(L\)) when \(\lambda \in (0, (\sqrt{2}-1)/L)\). Unlike the extragradient method, it requires only one projection per iteration.\\
\textbf{(v)~Golden Ratio Algorithm (GRAAL) \cite{malitsky2020golden}:}  
\begin{align*}
    y^k &= (1-\beta)u^k + \beta y^{k-1}, \\
    u^{k+1} &= \pi_{\mathcal{C}}(y^k - \lambda F(u^k)),
\end{align*}
with \(\lambda\) as the stepsize and \(\beta \in \left(0, (\sqrt{5}-1)/2\right]\). This method ensures convergence for Lipschitz operators (with constant \(L\)) when \(\lambda \in (0, 1/(2\beta L))\) and requires one projection per iteration. Additionally, the stepsize can be chosen adaptively, leading to the Adaptive Golden Ratio Algorithm (aGRAAL) \cite{malitsky2020golden}: 
\begin{align*}
    \lambda_k &= \min\left\{(\beta + \beta^2)\lambda_{k-1}, \frac{\|u^k - u^{k-1}\|^2}{4\beta^2\lambda_{k-2}\|F(u^k) - F(u^{k-1})\|^2}\right\}.
\end{align*}
\textbf{(vi)~{Operator splitting methods (DR)} \cite{facchinei2003finite}:}\\
In these types of methods, the operator \( F \) can be split into a summation of different operators. Then, solving \(\mathrm{VI}(\mathcal{C}, F)\) is equivalent to solving \(0 \in F_1 + F_2\), where \(F + \mathcal{N} = F_1 + F_2\), and the iterative update of each sub-operator leads to convergence towards the solution. A famous class of these algorithms is the Douglas-Rachford splitting method, which can be written as:
\begin{align*}
   y^{k+1} &= (I + \lambda F_2)^{-1}\left(u^k - \lambda F_1(u^k)\right), \\
   u^{k+1} &= u^k + \gamma(y^{k+1} - u^k),
\end{align*}
where $\gamma \in (0,2)$ is a relaxation parameter. The convergence of this method is guaranteed for different cases of $F_1$ and $F_2$; we refer interested readers to~\cite{giselsson2017tight} for further details.

{Recently, the authors in \cite{baghbadorani2025frank} proposed a projection-free algorithm for solving strongly monotone VIs. Instead of performing potentially expensive projections, the method uses a linear minimization oracle, which can tremendously reduce the computational complexity.} We also refer interested readers to \cite{mignoni2025monviso} for additional algorithms and applications of monotone variational inequalities, as well as a ready-to-use Python package.
\bibliography{ifacconf_ref}             
                                                   






\end{document}

%% file: block_closed_GNE.tex
\begin{tikzpicture}[>=latex, node distance=1cm, auto]

    \node [draw, block] (plant) at (0,0) {$x^+ = Ax + \tsum_{i\in\mc I} B_i u_i$};
    \node [draw, block, below=.3cmof plant] (GNEP) { $\begin{aligned}
        &~~\bu^* \text{ that solves \eqref{eq:def_P2}}
    \end{aligned}$ };

    \node [draw=none, left=of plant] (w) {};
    \node [fill=black, circle, right=of plant, inner sep=1pt] (dot) {};

    \node [draw=none, right=of dot] (x) {};


    \draw[-] (plant.east) --  node[above] {$x$}  (dot.west);
    \draw[-] (dot.south) |- (GNEP.east);
    \draw[->] (dot.east) -- (x.west);
    \draw[-] (GNEP.west) -| (w.east);
    \draw[->] (w.east) -- node[above] {$\bu^*[0]$}  (plant.west);
      
\end{tikzpicture}